# Study of electrostatic septum design and its high-voltage discharge protection

Zi-Feng He [a,*], Lian-Hua Ouyang [a,†], Man-Zhou Zhang [a], De-Ming Li [a], Zhi-Ling Chen [a],
Xiao-Bing Wu [b], Yue-Hu Pu [a,b]

[a] Shanghai Institute of Applied Physics, Chinese Academy of Sciences, Shanghai 201800, China
[b] Shanghai APACTRON Particle Equipment Co. Ltd, Shanghai 201800, China

## Abstract:

In this paper, we introduce the design of electrostatic septum (ESS) for the accelerator of Shanghai Advanced Proton Therapy (SAPT), and discuss its mechanical structure and the material selection of the electrode. The beam loss on the septum is studied, and the calculation results are given by the particle simulation and by the formula which is related to the placement angle and the divergence angle in the horizontal direction. Considering the thermal effect of the beam loss on the head of the septum, the equilibrium temperature at work is calculated. In addition, the distribution of the electric field and the trajectory of the particles are also simulated. The phenomenon of vacuum discharge in the working ESS is analyzed in detail on the relationship between the working current and the enhancement factor of the electric field on the electrode surface. Based on the discharge mechanism, the importance of degassing and cleaning of ESS is analyzed. We also analyzed the effect of external series resistance from circuit and discharge mechanism. It is considered that series resistance within a certain range can increase the breakdown voltage between vacuum electrodes and reduce the irreversible damage to the surface of the electrode system caused by the discharge process. The formula for the value of the resistance is given.

## Key words:

electrostatic septum, vacuum discharge, high voltage condition, discharge protection, proton medical accelerator

## I. Introduction

High-voltage electrostatic septum (ESS) is one of the most critical elements used in the beam injection process or the beam extraction process in the accelerator of Shanghai Advanced Proton

---

[*] hezifeng@sinap.ac.cn
[†] ouyanglianhua@sinap.ac.cn

Therapy (SAPT) and other similar multi-stage accelerator systems.[1] As a medical proton accelerator for daily treatment, it demands very high reliability, availability and serviceability. SAPT utilizes multi-turn injection and third order resonant slow extraction method. Between the septum and the electrode, there is high voltage which generate the electric field for a certain deflection angle of beam passing through the plates. Then the beam goes in or extact out of the ring accelerator under the matching of other magnetic components.

In order to eliminate the phenomenon of vacuum discharge and to avoid high voltage breakdown, the electric field between the electrostatic deflection plates is limited. Nevertheless, as a high voltage device, the sparking occurs occasionally between the electrodes, due to the increase of the temperature of the septum and the increase of the beam loss level in the ring channel. Then the voltage of the electrostatic septum drops suddenly. The accidental discharge can cause injection difficulties, beam trajectory removal or extraction instability. If not treat in time or without a certain discharge withstand capability, the high voltage components may even be damaged during frequent discharge or intense discharge.

Therefore, in the accelerator design, the high voltage requirements of the equipment is need to be compatible with the requirement of high voltage insulation. Some methods, such as to optimize the structure and the material selection, to maintain the cleanliness of the surface of the device and to monitor the external circuit, are necessary to minimize the occurrence rate of the discharge phenomenon as much as possible to ensure the beam quality of the accelerator. It is shown in some literatures that different equipment builders have given their own solutions, and experimental researchers have also obtained some interesting but inconsistent results. In this paper, we will try to give a quantitative analysis result in combination with the vacuum discharge mechanism, the transient analysis of the circuit, and the thermal effect of the material under pulse power.

## II. Structural design

The electrostatic septum is a plate-shaped electrode used in a vacuum environment, as shown in Fig. 1. There is a certain number of curvatures of in the electrode of in the injection system, while the electrodes of the extraction system are flat. The septum is connected to the earth, and the negative electrode is connected to the high voltage power supply. The structure of the septum can be designed as different shapes, such as the sheet[2,3], the ribbon[4-6], the wire[7,8] or the mixed by wire and sheet[9], which are fixed on a C-type of support yoke. There is no electric field in the channel between the septum and the support yoke, and the circle beam can pass through it. On the other hand, the channel between the septum and the high-voltage electrode has strong electric field and can pass through the injected or extracted beam with a certain deflection angle.

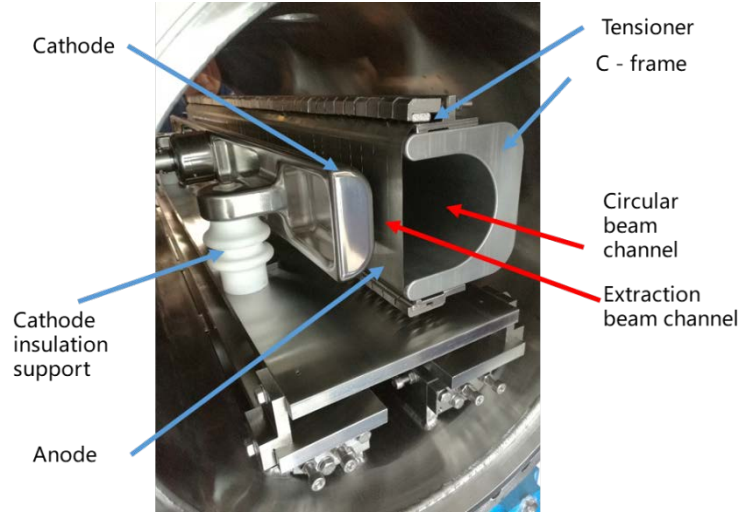

FIG. 1. Electrostatic septum for the beam extraction system in SAPT.

## A. Selection of the electrode materials

The purpose of electrode material selection is to enable the electrode to maintain a uniform electric field distribution, and to resist the permanent damage on the surface of the electrode during discharge. The microscopic deformation of the electrode surface will cause electric field distortion when it is bombarded by particles. It can suppress the field emission of electrons by selecting the proper material considering the melting point and the thermal conductivity. The commonly used electrode materials are shown in Table I.

TABLE I. Electrode materials and their physical properties.

| Material | Atomic Number | Work Function [eV] | Melting Points [K] | Specific Heat [J/(kg·K)] | Thermal Conductivity [W/(m·K)] | Density [g/cm$^3$] | Resisivity [nΩ·m] (at 20 °C) |
|---|---|---|---|---|---|---|---|
| **Al** | 13 | 4.06–4.26 | 933.47 | 897 | 237 | 2.70 | 26.5 |
| **Ti** | 22 | 4.33 | 1941 | 523 | 21.9 | 4.506 | 420 |
| **Cu** | 29 | 4.53–5.10 | 1357.77 | 385 | 401 | 8.96 | 16.78 |
| **Mo** | 42 | 4.36–4.95 | 2896 | 251 | 138 | 10.28 | 53.4 |
| **Ta** | 73 | 4.00–4.80 | 3290 | 140 | 57.5 | 16.69 | 131 |
| **W** | 74 | 4.32–5.22 | 3695 | 134 | 173 | 19.25 | 52.8 |
| **Re** | 75 | 4.72 | 3459 | 137 | 48.0 | 21.02 | 193 |
| **SUS304** | – | 4.4 | 1425 | 522 | 15 | 7.93 | 72 |

The anode materials should have high melting point and high thermal conductivity to reduce the probability of local evaporation during the process of high voltage breakdown. The design using a whole piece of plate, such as tantalum plate, is of simple structure and broken difficultly. But the deformation after high temperature breaking makes a bad effect on the effective thickness. The design using wires, such as Rhenium-tungsten alloy with a diameter of 0.05–0.08 mm, is easy to make a smaller thickness septum. However, the disadvantage is that the wires on the inlet-end have the risk of fracture which causing high voltage flash. For the design using ribbons made of molybdenum, it is a compromise choice of no easy to fracture and small deformation, but it requires high accuracy for assembly.

The main purpose of cathode material selection and its processing is to reduce the dark current. The cathode is connected to a negative high voltage power supply and is suggested to made of light element which has a high work function and a low conductivity. Through the experiments on cathodes made of different materials, it is found that titanium and titanium alloy (TiAl$_6$V$_4$) have the best effect after nitriding, copper and stainless steel can achieve better effect after degassing, while stainless steel without degassing or titanium after oxygen treatment has larger dark current that should be reduced. Some literature mentionss that it is a good choice to use stainless steel (SUS04) or aluminum after surface oxidation treatment.[10,11] We use the stainless steel cathode with finishing and degassing.

## B. Molybdenum sheet and beam loss

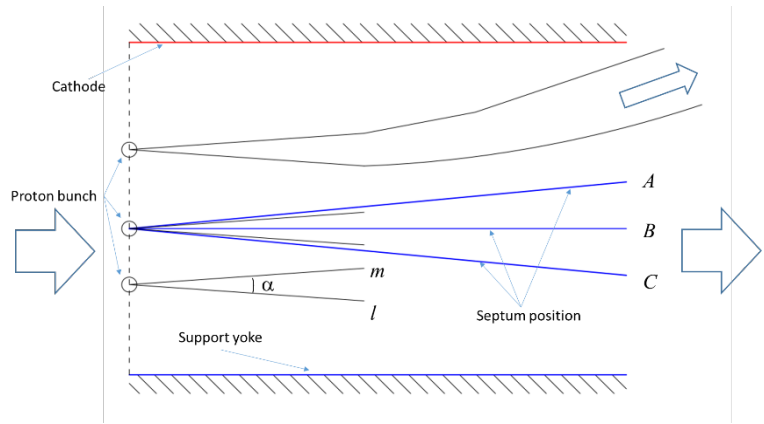

FIG. 2.  Beam loss on the molybdenum sheet

The placement of the septum which is in the proton beam motion region causes a significant beam loss on the septum surface. The beam loss here produces intense X-ray, γ-ray and neutron. These high energy rays can cause radiation damage to the components.[12] The effective thickness $T_e$ of the septum is the key parameter, which is determined by the following formula,[13]

$$T_e = T_p + T_g + T_s \tag{1}$$

where $T_p$ is the physical thickness of the septum; $T_g$ is the thickness caused by geometric alignment or installation error and is about 50 μm with reference to the general installation accuracy; $T_s$ is the thickness of the shadow, that is, considering the particle hitting the side of the septum. It is difficult to fully satisfy the Hardt condition at the entrance of the electrostatic septum. When the beam passes through with a certain emittance, it is inevitable that there will be collisions between the particle and the side wall, as shown in Fig. 2. In the case of a certain emittance, the inclination angle x' of the particle at each point in the X direction of the incident plane is contained between the two sides of $l$ and $m$, where the angle is α about several milli-radians between them. Taking $l$ as the reference direction, the intersection angle of the septum and $l$ is θ, then Ts can be expressed as show in Eq. (2),

$$T_s = \begin{cases} \dfrac{1}{2}\left[v(\alpha-\theta)\right]^2 \dfrac{m_p}{q_p E\sqrt{1-\beta^2}}, & \theta < 0 \\ L\theta + \dfrac{1}{2}\left[v(\alpha-\theta)\right]^2 \dfrac{m_p}{q_p E\sqrt{1-\beta^2}}, & 0 \le \theta \le \alpha \\ L\theta, & \theta > \alpha \end{cases} \quad (2)$$

where $v$ is the velocity of the proton, $\beta = v/c$, $E$ is the electric field of the septum, $m_p$ and $q_p$ are the rest mass and charge of proton respectively.

$T_s$ is the shadow thickness, that is, considering the particles hitting the side of the septum, which is due to the emittance of the beam leading to the collision between the particles and the side of the septum. The relationship between effective thickness Te and beam loss η is as shown in Eq. (3), [14]

$$\eta = \dfrac{fT_e}{s} \quad (3)$$

where s is the spiral step, $f$ is the horizontal direction change caused by the nolinear growth of the β function and is bout 1.5.

For 250 MeV proton beams, the smallest $T_s$ is about 37 μm after 60 kV/cm between the electrodes in the case of $\alpha = 1$ mrad. The length of the corresponding side wall, that is $Ts/\alpha$, is about 37 mm. So the beam loss on the septum is mainly distributed in the area of 37 mm along the wall from the entrance end. The effective thickness $T_e$ is 190 μm for the 100 μm septum considering the installation error and shadow thickness. Then the minimum beam loss is 1.9% when the spiral pitch s = 1.5 cm.

The beam loss is related to the location angle of the septum[15] as shown in Fig. 3, where the $\theta$ is the angle between the septum and the direction $m$. The septum should be installed in the position that the angle with the direction $m$ is avoided to be greater than zero. The beam loss on the septum can be kept at a low level by adjusting or working in the angle region where the $\theta$ is slightly less than zero. That is to say, in the optimal case, the installation angle of the septum is as much as possible along the direction $m$. So that the beam loss on the side wall without field is close to zero, while the beam loss on the side wall with field is the smallest. The calculated resulted are consistent with which obtained by particle tracking method.[15]

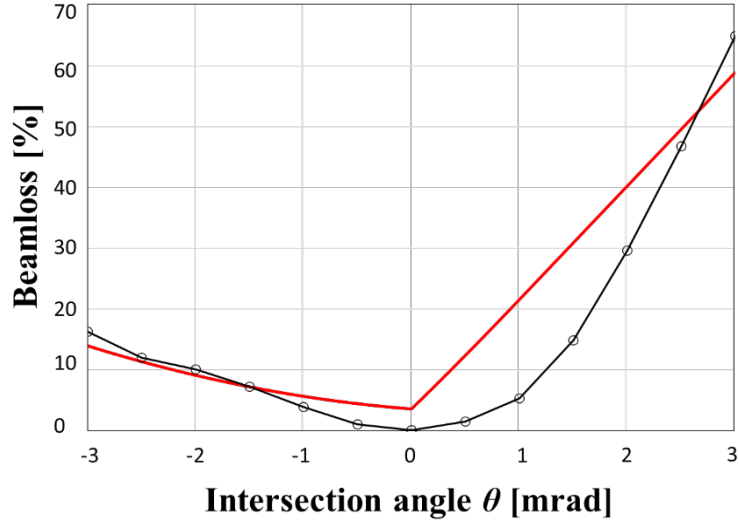

FIG. 3. Relationship between beam loss and installation angle of the septum. The red line is from the Equa. (2), and the black line is from simulation result.

## C. Thermal effect calculation

### 1. Beam loss on the sheet

The temperature of the septum is estimated as follows: the number of particles in the synchrotron is $8.35\times10^{12}$, with the energy of 250 MeV, $\beta$ value of 0.61361 and the revolution frequency of 7.478 MHz. The total energy of beam is 33334.45 J. It is supposed that 20% of the beam, that is 66.89J, is lost in the slow extraction process especially at the head of the septum. The power density on the septum surface is $2.09\times10^5$ W/m$^2$, considering the area of 2 mm × 40 mm and the injection period of 4 second.

### 2. Equilibrium temperature on the sheet

According to the above power density, the calculation shows that the septum can reach a balance temperature of 650 °C after 50 seconds, as shown in Fig. 4. The calculation also shows that the thermal radiation in vacuum plays an important role in the heat dissipation. The hightest temperature of the septum can reach 950 °C if only considering the heat conduction process. The working temperature is lower than that calculated in the case from Olivo[13] and Goddard[8].

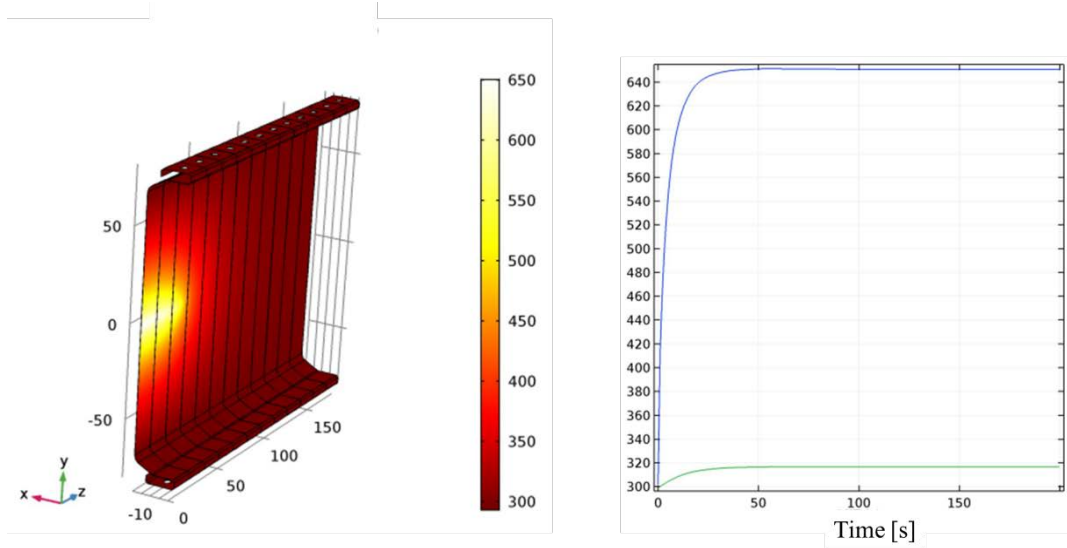

FIG. 4. The temperature distribution and time distribution of the septum, taking into account the radiation heat dissipation in the vacuum.

## III. Simulation for the electric field distribution and trajectory

In order to reduce the beam loss in the magnet, it is necessary to ensure the beam at the entrance has a certain separation distance form the circulating beam and a certain deflection angle. The relationship between the deflecton distance $\Delta x$ and the deflection angle $\theta$ of the particle is shown in Eq. (4),[1]

$$\Delta x = \theta \sqrt{\beta_{ES} \beta_{MS}} \sin \mu, \tag{4}$$

where the $\beta_{ES}$ and $\beta_{ES}$ are the envelope function values for the electrostatic septum and the magnetostatics deflection magnetic respectively, and $\mu$ is phase difference of the two components generally selected as 90°+n×180°. The particle with the mass of $m_p$ and the charge of $q$ will be deflected by the angle $\theta$ in the electric field $E$, as shown in Eq. (5),[16]

$$\theta = \arctan\left(\frac{qEL}{pc\beta}\right), \tag{5}$$

where $L$ is the length of the septum, $E$ is the electric field, $q$ is the charge of the particle, $p$ is the particle momentum, c is the speed of light and $\beta$ is the relativistic normalized speed. The parameters for extraction and injection are given in Table II. The electric field distribution is shown in Fig. 5, and the simulation of particle trajectories is shown in Fig.6.

|  | Energy (Ek) | Deflection angle ($\theta$) | Septum length (L) | Electric field (E) | High Voltage (V) |
|---|---|---|---|---|---|
| **Injection** | 7 MeV | 185 mad | 0.5 m | 52.2 kV/cm | 62.65 kV |
| **Extraction** | 250 MeV | 5 mrad | 1.0 m | 60.4 kV/cm | 60.4 kV |

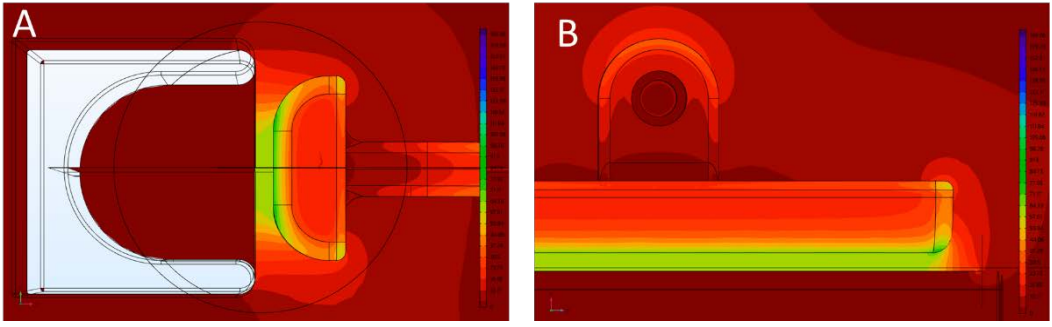

FIG. 5     The distribution of the electric field, (A) on the mid-surface, (B) on the horizontal plane.

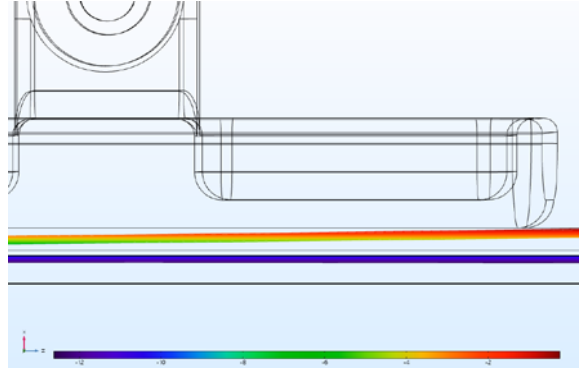

FIG. 6     The particle trajectory simulation, where the color represents the X direction coordinates of particles.

## IV. discharge phenomena

The discharge of the septum is a breakdown phenomenon of the electrode in the vacuum. When the distance between the electrodes is within the range of $10^{-4} \sim 10$ mm, the corresponding electric field strength is constant within the range of measurement error.[17] The joule heating mechanism considers the joule heat by the electrode emission is relate to the current density. The criterion for breakdown in steady state is $J = \pi h^{-1}(\lambda/k_0)^{1/2}/2$, where h is the electrode microprojection, $\lambda$ is the thermal conductivity of the electrode, and $k_0$ is the temperature coefficient of the resistivity k of the material that $k = k_0 T$. This means that the selection of the electrode materials with high thermal conductivity and low temperature coefficient is helpful to raise the criterion value and then suppress the discharge phenomenon.

The electron emission mechanism, which plays a leading role on the septum, is field emission. It is need to considered two field enhancement factors. One is $\beta_0 \approx 10$ which is due to the geometrical nonuniformity of the electrode surface mainly caused by micro-bumps, another is $\beta_1 \approx 10\text{--}10^2$ which is due to microscopic inhomogeneity of electrophysical properties mainly caused by micro-additions on the surface. The total field enhancement factor is $\beta = \beta_0\beta_1$,[18] which should be noted that it will decrease after conditioning procession.

According to the Fowler-Nordheim equation, the field emission current density J [A/cm2] of the metal cathod is calculated as Eq. (6),[19]

$$J = \frac{AE^2}{\phi t^2(y)} \exp\left[\frac{-B\phi^{3/2}}{E} v(y)\right], \qquad (6)$$

where A = 4.541×10⁻⁶, B = 6.831 ×10⁻⁷, y = 3.795 ×10⁻⁴$E^{1/2}\phi^{-1}$. In the Eq. (6), $t(y)$ and $v(y)$ can be approximately calculated as t(y) = 1+0.1107$y^{1.33}$, v(y)=1-$y^{1.69}$, respectively.[20] E [V/cm] is calculated using the factor β, such as E = β$E_0$, where $E_0$=U/d is the electric field produced at the macroscopic dimension d with the voltage U.

The key to suppress the initial electron emission is the electrode treatment. The result of field emission current is shown in the following chart, in the case that the voltage is 75 kV, the emission area is 100cm×cm and the work function is of stainless steel. Our system, without special cleaning process, can be maintained when the leakage current is less than 10×10⁻⁶ A, but it is very easy to enter the breakdown state when the leakage current is large than 60×10⁻⁶ A. The reason is that some processes will increase the field enhancement factor and lead to increase the emission current under the long-term high voltage action. These processes are depending on the metal species, surface contamination, temperature and electric field, such as surface diffusion, evaporation, adsorption and contamination migration. The emission current should be less than 0.1×10⁻⁶[A] after thorough conditioning for the systems that are well assembled and clean enough. It is critical important to perform special cleaning after processing about the electrode for reducing dark current.

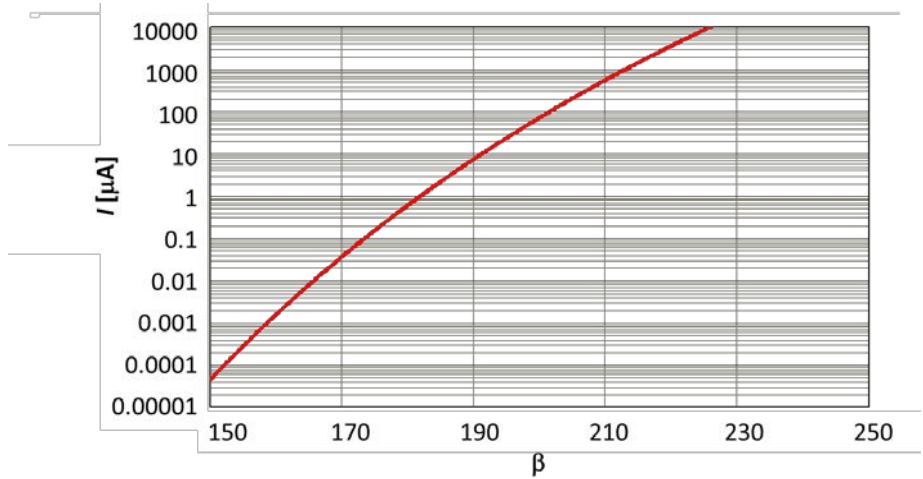

FIG. 7   The relationship between emission current and the field enhancement factor

The relationship between breakdown voltage *U* and electrode gap *d* is $U = Cd^\alpha$, where C is constant. For steady-state high voltage, the coefficient α has different values according to *d*. In the case of $d \leqslant$ 1 mm, α = 1, then C is the electric field. In the case of *d* > 1 mm, it is usually that α < 1. The empirical formulas include $U = 97d$ for d < 0.4 mm, $U = 58d^{0.58}$ for 0.4 mm < d < 40 mm and $U = 123d^{0.34}$ for 40 mm < d < 100 mm.

The ESS is a parallel electrode system with an electrode area of 800 cm² and gap of 10 mm. The maximum voltage that the electrode can withstand is about 220 kV. This system is usually possible to achieve long-term stable operation bellow 100 kV, if it is well assembled, with a high degree of cleanliness and with a good vacuum, since these methods can significantly suppress the increase of the field enhancement factor on the surface. Referring to the calculation result of Fig. 8, the emission current from the electrode will be reduced, thereby improving the stability of the electrode with high voltage.

## V. Surface treatment and high-voltage conditioning

As can be seen from Fig. 8, the working current of the ESS, usually generated by field emission, is greatly affected by the field enhancement factor on the electrode surface. It is necessary to do surface treatment on the surface to reduce the field emission current and to ensure the field enhancement factor does not change greatly during the operation.

In the processing and manufacturing of the electrode, it is necessary to ensure the flatness of the surface, to reduce the microscopic protrusion and to form a mirror surface of the electrode. Electron beam surface treatment (EBEST) method, that surface impurities and protrusions form a smooth remelting layer by electron beam, is proved to work well.[21] As for chemical and electrochemical polishing methods, they can improve the surface roughness and cleanliness, but they provide limited improvement for vacuum insulation.[22]

No-metallic addenda and dust particles on the surface of the electrode are always inevitable. Some are formed on the surface during machining, others may fall to the surface during vacuum or operation. It is very important for improving the performance of ESS to clean the electrode surface and even the inner vacuum interface. In operation, the internal components need to be soaked in acetone, washed with deionized water at room temperature, ultrasonically cleaned with alcohol and dried with dry clean gas finally. The cleaning process is very effective in removing internal contamination and oil film, greatly reducing the particles and dust generated from vacuum components in high voltage environment, greatly improving the performance of vacuum insulation.

The adsorbed gas on the electrode surface is also an important factor affecting vacuum insulation.[23,24] At high voltage, the plasma formed by these adsorbed gases has a significant negative impact on vacuum breakdown. This is because the complete breakdown is usually accompanied by the deterioration of vacuum. The vacuum will become worse to a certain value which will lead to the intense discharge between the electrodes. Therefore, it is very important to reduce the adsorption gas inside the vacuum system. After processing and cleaning, the surface of components in the vacuum chamber still contain various gas molecules adsorbed in the atmosphere. These gas molecules can be removed by vacuum baking. In practice, it is better to use nitrogen protection than to expose the atmosphere directly when destroy the vacuum after degassing by vacuum baking. In this way, both the maintenance of voltage insulation level and speed of the high voltage condition are better.

As shown in Fig. 6. below, the ESS, which has been fully cleaned and vacuum baked, performs very well in the high voltage conditioning compare with the previous state. The curve of Ip is working current of one ion pump while being conditioning. In practice, it is found that the curve of Ip has a good correlation with the curve of vacuum P during the increase of the voltage V. The value of Ip is a very obvious indicator for the change of the vacuum system, for it can react more sentively to the outgassing and degassing state in the vacuum chamber during conditioning. Before the treatment, it needs more than 45 hours before the system reached 53.4 kV. On average, the increase is only 0.675 kV per hour. But after special treatment, the system can reach the high level of 75 kV within 5 hours. After the voltage dropped form 75 kV to 65 kV, the system was kept for 140 minutes without any sparking. This means that the vacuum insulation state of the ESS fully meets the working needs.

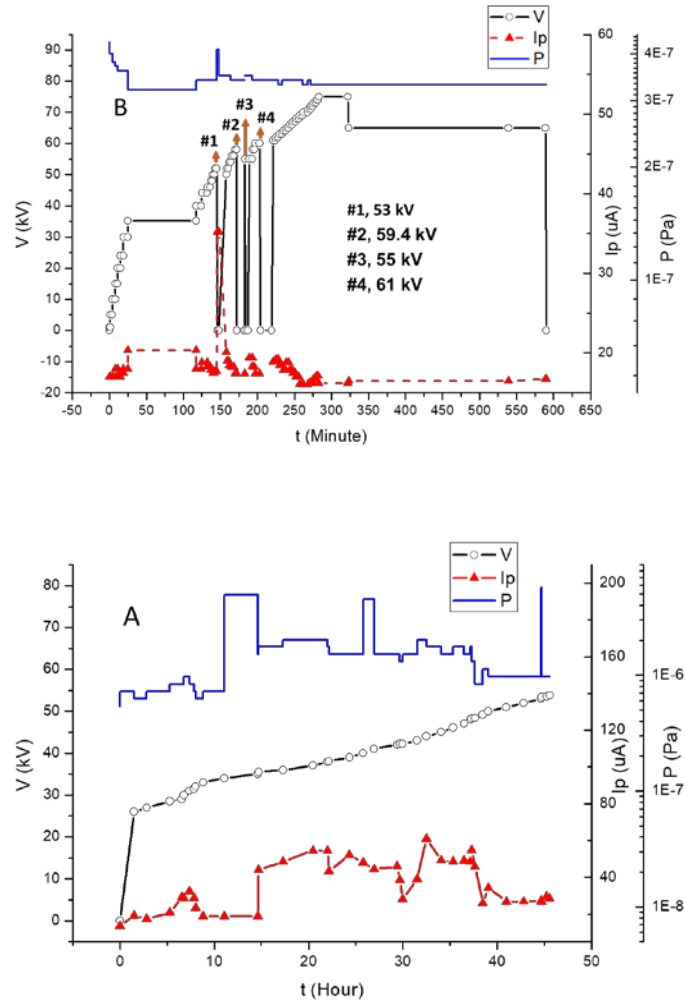

FIG. 8  High voltage conditioning for the ESS

A. Without baked and clean. B. After baked and clean. V is the voltage between the electrodes, Ip is the current of the ion pump being working, and P is the value of the vessel.

## VI. Influence of external circuit to the high-voltage electrode

### A. Increase the breakdown voltage of the vacuum electrode

Hackam did a experiment about dependence between external resistance $R_d$ through series resistor and breakdown voltage $U_b$, which shows that $U_b$ can be increase when the $R_d$ is increased in the range of $10^{-1}$–$10^4\,\Omega$, but $U_b$ does not change when the value of Rd is more than 10 kΩ.[25,26] Whether it is DC or AC high voltage, such phenomenon will occur in different sizes of gaps. From the circuit point of view, the series resistance is related to the time constant $RC$, which determines the discharge speed of discharge current in the circuit. Increasing the resistance will increase the value of $RC$, that is, the discharge speed of current will slow down. For the system with pF capacitance and the resistance of $10^3$ Ω, its time constant is of nanosecond correspondingly. The influence of this magnitude on the breakdown voltage is very obvious, which is related to the time characteristic of arc growth during discharge in vacuum.

According the discharge current analysis from Jüttner, it takes tens of nanoseconds that the arc initially develops to its maximum.[27] For an electrode with gap $d$, capacitance value of $C$, and initial voltage of $U_0$, variation curve of the discharge current I and the voltage U with time t during discharge can be obtained by the following equation,[28]

$$U(t) = U_0 f(t), \tag{7}$$

$$I(t) = A U_0^{3/2} v_c t (d - v_c t)^{-1} f(t)^{3/2}, \tag{8}$$

$$f(t) = \{1 - 0.5B[\ln(1 - v_c t / d) + v_c t / d]\}^{-2}, \tag{9}$$

$$B = U_0^{1/2} A d / C v_c, \tag{10}$$

where the constant A is $3\times10^{-5}$ AV$^{-3/2}$, and $v_c$ is the plasma velocity in the breakdown stage. The ESS is a high voltage system with cathode of stainless steel and anode of molybdenum, which is worked with a maximum voltage of 80 kV. The curves of current and voltage as a function of time during discharge-breakdown is calculated as follows, considering $v_c$ = 26 km/s, d = 10 mm, C = 71 pF.

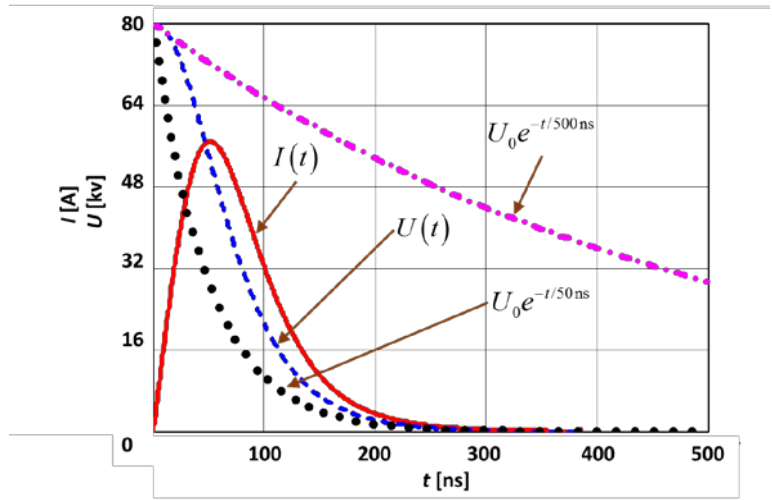

FIG. 9. Voltage and current of ESS during discharge-breakdown state. *I(t)* is the relationship between arc current and time, *U(t)* is the relationship between the voltage of electrodes and the time, $U_0 e^{-t/50ns}$ is the relationship between electrode voltage and time when the time constant of circuit is 50 ns, and $U_0 e^{-t/500ns}$ is the relationship between electrode voltage and time when the time constant of circuit is 500 ns.

In the above figure, when the arc current I(t) reaches the maximum of 56.88A, the time is tc= 51 ns. However, when the circuit is connected in series with different resistors, the time constant of the circuit can be adjusted correspondingly to slow down the energy dissipation of the electrode. Taking the time constants 50ns and 500 ns as example, the corresponding series resistance values are 704 Ω and 7 kΩ. At the time of 51 ns, the residual voltage of the electrode are 28.85 kV and 72.24 kV correspondingly, which means that the system energy $CU^2$ remains 13% or 82% respectively. This indicates that series resistance can slow down the current growth at the beginning of the discharge. In order to make discharge between the two electrodes, it is necessary to further increase the voltage, that is, series resistance can increase the discharge voltage. This result is consistent with the research conclusion of Hackam and Jüttner. Therefore, in order to increase the discharge voltage of the system, the series resistance $R_d$ of the electrode should have the following relationship with the electrode capacitance $C$,

$$R_d \geq 10 \frac{t_c}{C} . \tag{11}$$

Martinis told that the resistance in the high voltage circuit will not change the breakdown voltage of the electrode system.[29] However, we do not think this conclusion contradicts the results from Hackam[25], for the range of the resistance value in Martinis's paper is 1 MΩ – 30 MΩ. From the previous analysis, it can be seen that the breakdown voltage will not increase when the series resistance of the electrodes is greater than $10^4$ Ω, while the breakdown voltage will decrease obviously with the decrease of the resistance when the resistance is less than $10^4$ Ω. That is the slow-down effect of the resistance on the growth of arc current.

## B. Reduce the irreversible damage to the surface

Martinis believes that the series resistance is to reduce the irreversible damage to the surface of the electrodes and insulators during high-voltage breakdown, which is helpful to the stability of ESS.[29] It is usually to series the resistance between the high voltage power supply and the ESS device when the ESS is under conditioning or in the actual operation. The storage energy E of the ESS is determined by the working voltage. Whether the storage energy E gives the irreversible damage to the electrode surface is related to the resistance Rd. We believe that the series resistance can reduce the arc power of the bombardment on the metal electrode, reduce the temperature rise of the bombardment center, and then prevent the central area of the bombardment from melting. In this way, the electrodes are protected.

The temperature rise $\Delta T$ can be calculated by the following formula[30,31],

$$\Delta T = \frac{E}{A} \frac{1}{\sqrt{\tau}} \sqrt{\frac{4}{\pi c_p k \rho}} , \tag{12}$$

where $E = CU^2/2$ is the storage energy of ESS with capacitance $C$ and voltage $U$, $A$ is the area where the electrode is bombarded, $c_p$ is specific heat, $k$ is the thermal conductivity, $\rho$ is the density and $\tau$ is the interaction time between the arc and the electrode. The value of $\tau$ can be determined as the time constant of discharge circuit,

$$\tau = R_d C . \tag{13}$$

After a period of time constant, the storage energy will be released by 86.5%. The relationship between temperature rise $\Delta T$ and series resistance $R_d$ is derived from the above two formulas, as shown below,

$$\Delta T = \frac{U^2}{A} \sqrt{\frac{C}{R_d}} \sqrt{\frac{1}{\pi c_p k \rho}} . \tag{14}$$

So, in order to make the temperature not exceed the melting point or maximum working temperature $T_m$, $R_d$ should also meet the following condition,

$$R_\mathrm{d} \geq \frac{CU^4}{T_\mathrm{m}^2 A^2} \frac{1}{\pi c_\mathrm{p} k \rho} \quad . \tag{15}$$

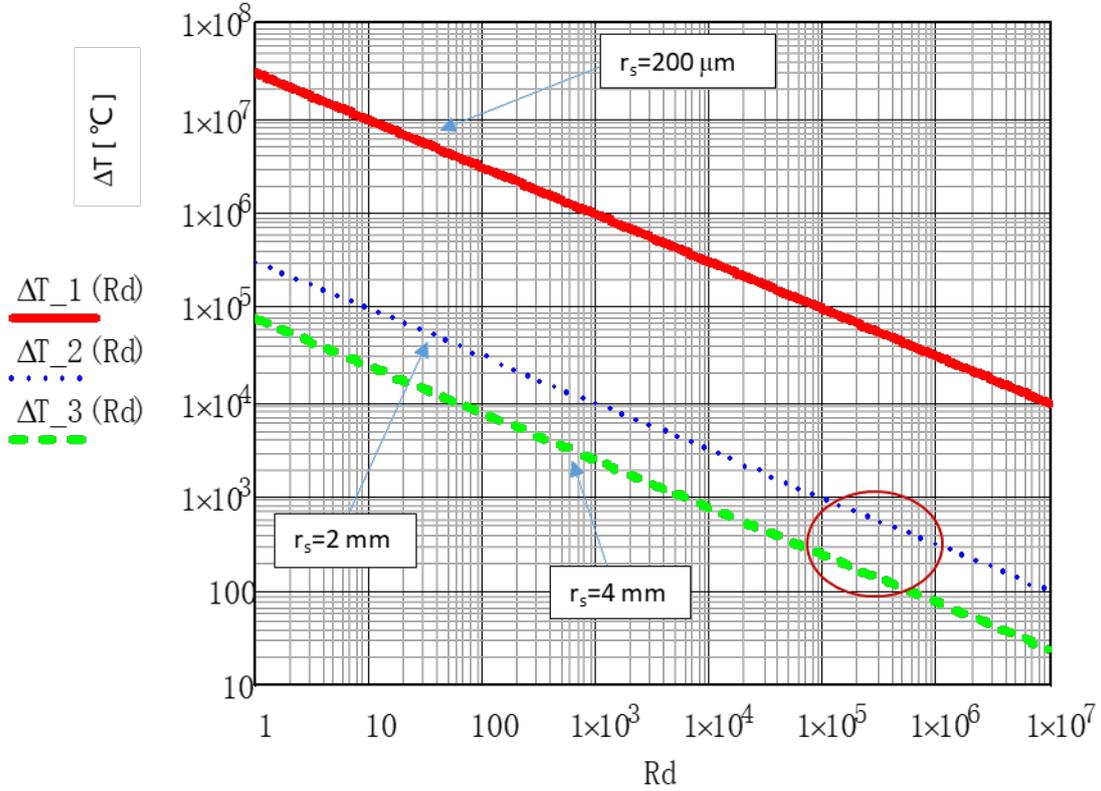

FIG. 10.  The relationship between temperature rise $\Delta T$ and series resistance $R_\mathrm{d}$ in different cases that rs = 200 μm, rs = 2 mm, and rs = 4 mm, where rs is the radius of the bombard area.

We can see that there are many obvious discharge marks on the anode surface, which are the spots with radius about 2 mm. According to the above calculation, the series resistance near $10^5$ Ω should be used to limit the temperature rise to $10^3$ ℃, which is help to decrease the damage to surface, and then make sure the stability of the ESS.

TABLE II.  Resistance in the high voltage circuit of ESS

|  | **Accelerator type** | **Voltage** | **Resistance** |
|---|---|---|---|
| S. I. N. Swiss, 1975[13] | Proton, cyclotron | 80 kV | 25 MΩ+700 Ω |
| Univ. of Milan, Italy, 1984[32] | Proton, Superconducting Cyclotron | 120 kV | 1 MΩ |
| Univ. of Milan, Italy, 1986[29] | Proton, Superconducting Cyclotron | 120 kV | 1 MΩ+650 Ω |
| FAIR, GSI, Germany，2007[21] | Proton/Ion, synchrotron | 300 kV | 5 MΩ+1 kΩ |
| MedAustron, Austria, 2013[3] | Carbon, synchrotron | 80 kV | 150 kΩ |

## C. Value of resistance

The required minimum resistance can be obtained from the Equations (11) and (15). It should be noted that the surge current will generate appreciable voltage drops on the series resistor. When the voltage drop exceeds its insulation level, the resistor will be bypassed and then cannot provide the above protection. Therefore, the resistor is usually insulated with oil or placed in the vacuum institutions. On the other hand, the value of Rd should also be limited so that the voltage drop does not exceed its insulation level.

# VII. Conclusion

The ESS is the key device for the particle beam injection and extraction of the medical proton synchrotron, and it more frequently fails in the actual accelerator commission than other devices. The cause of the fault is mainly the failure of high voltage insulation, such as the accidental discharge of this kind of equipment is difficult to completely avoid. Due to the special of high-voltage equipment, it is particularly necessary to ensure the high requirement of the stability during working, especially to ensure the safety and reliability while the therapy treatment is processing.

The beam loss mainly occurs on the septum at the inlet end, with the resonant slow extraction method. It can reduce the beam loss and optimize to the best level by controlling the place angle of the septum and by minimizing the divergence angle when the beam enters the ESS. Reducing the beam loss can simultaneously reduce the temperature rise and radiation intensity generated on the head of the septum. The particle trajectory and electric field calculation show that the design of the ESS meets the physical requirements of SAPT.

The calculation of the vacuum discharge current shows that the leakage current of the high voltage power supply will change significantly with the change of surface enhancement factor. The change of the enhancement factor, which is dynamic while the ESS is working, will lead to discharge failure if it is not suppressed. The conditioning process in our situation also shows that improving this enhancement factor requires surface treatment, cleaning and degassing for the internal electrodes. Strict requirements in these steps are particularly critical for the ESS finally has a high level of high voltage performance.

Using Jüttner's study of discharge current, we theoretically explain the experimental phenomena from Hackam and Martinis and derive the resistance discriminant which helps to improve the breakdown voltage of the vacuum electrode. Based on the instantaneous effect of the discharge current on the surface of the electrode, it is theoretically explained that the significance of external resistance in preventing the surface from being irreversible damage. We also give the formula for the value of this resistor.

# Acknowledgments

This work was supported by National Key Research and Development Plan of China [2016YFC0105400].